\documentclass[journal]{IEEEtran}
\ifCLASSINFOpdf
\else
\fi
\usepackage{algorithm}
\usepackage{algpseudocode}
\usepackage{graphicx}
\usepackage[justification=centering]{caption}
\usepackage{dblfloatfix}
\usepackage{multirow}
\usepackage{etoolbox}
\usepackage{array}
\usepackage{subcaption}
\usepackage{ amssymb }
\usepackage[fleqn]{amsmath}
\usepackage{multirow}
\usepackage{placeins}
\usepackage{comment}
\usepackage{subcaption}
\usepackage{placeins}
\begin{document}
\title{Fuzzy Unique Image Transformation: Defense Against Adversarial Attacks On Deep COVID-19 Models }

\author{Achyut Mani Tripathi,~\IEEEmembership{IIT Guwahati}
and  Ashish Mishra~\IEEEmembership{IIT Madras} 
         
        }
        %and~Jane~Doe,~\IEEEmembership{Life~Fellow,~IEEE}% <-this % stops a space
%\thanks{M. Shell was with the Department
%of Electrical and Computer Engineering, Georgia Institute of Technology, Atlanta,


\markboth{Journal of \LaTeX\ Class Files,~Vol.~14, No.~8, August~2015}%
{Shell \MakeLowercase{\textit{et al.}}: Bare Demo of IEEEtran.cls for IEEE Journals}



\maketitle

\begin{abstract}

Early identification of COVID-19 using a deep model trained on Chest X-Ray and CT  images has gained considerable attention from researchers to speed up the process of identification of active COVID-19 cases. These deep models act as an aid to hospitals that suffer from the unavailability of specialists or radiologists, specifically in remote areas. Various deep models have been proposed to detect the COVID-19 cases, but few works have been performed to prevent the deep models against adversarial attacks capable of fooling the deep model by using a small perturbation in image pixels. This paper presents an evaluation of the performance of deep COVID-19 models against adversarial attacks. Also, it proposes an efficient yet effective Fuzzy Unique Image Transformation (FUIT) technique that downsamples the image pixels into an interval. The images obtained after the FUIT transformation are further utilized for training the secure deep model that preserves high accuracy of the diagnosis of COVID-19 cases and provides reliable defense against the adversarial attacks. The experiments and results show the proposed model prevents the deep model against the six adversarial attacks and maintains high accuracy to classify the COVID-19 cases from the Chest X-Ray image and CT image Datasets. The results also recommend that a careful inspection is required before practically applying the deep models to diagnose the COVID-19 cases.

\begin{comment}
Early identification of COVID-19 using a deep model trained on chest X-ray and CT  images has gained considerable attention from researchers to speed up the process of identification of active COVID-19 cases. These deep models act as an aid to hospitals that suffer from the unavailability of specialists or radiologists, specifically in remote areas. Various deep models have been proposed to detect the COVID-19 cases, but very little work has been performed to prevent the deep models against adversarial attacks capable of fooling the deep model by using small perturbation in image pixels. This paper presents an evaluation of the performance of deep models against adversarial attacks and also proposes an efficient yet effective Fuzzy Unique Image Transformation (FUIT) technique that downsamples the image pixels into an interval. The images obtained after the FUIT transformation are further utilized to train the secure deep model that preserves high accuracy of the diagnosis of COVID-19 cases and provides reliable defense against the adversarial attacks. The experiments and results show the proposed model prevents the deep model against the six adversarial attacks and maintain the high accuracy to classify the COVID-19 cases from the Chest X-Ray Image and CT image Datasets. The results also recommend that a careful inspection is required before practically applying the deep models for the diagnosis of COVID-19 cases.
\end{comment}

\end{abstract}

\begin{IEEEkeywords}
Adversarial Attacks, Chest X-Ray, COVID-19, CT Image, Deep Models, Fuzzy Unique Image Transformation.
\end{IEEEkeywords}

% \fi
\IEEEpeerreviewmaketitle

\section{Introduction and Related Work}
The occurrence of a novel CORONAVIRUS \cite{wu2020new} challenges the healthcare systems of all across the world to control an exponential growth of CORONAVIRUS that first occurred in Wuhan and Hebei cities of the China \cite{wu2020new} in December 2019 and later spared to other countries across the world. Based on the degree of spread of the virus World Health Organization (WHO) declared the disease as COVID-19 pandemic \cite{collaborative2020global}. Cough, fatigue, fever, and illness in the lungs are among the earlier symptoms suggested by clinical experts for diagnosing COVID-19 cases at an initial stage. Control and prevention of the COVID-19 demand the maximum number of medical tests. Healthcare systems across the world suffer from a lack of effective testing toolkits to identify COVID-19 cases in a current situation. The early identification of COVID-19 cases would be helpful to quarantine the high-risk COVID-19 patients and also useful to break a chain of further spread of the virus in the community. 

\par In an attempt to develop a testing toolkit for the diagnosis of COVID-19, researchers from the radiology domain suggested the use of reverse transcription-polymerase chain reaction (RT-PCR) test \cite{ai2020correlation}. However, the test requires long latency to identify COVID-19 cases and demands highly expert radiologists \cite{ai2020correlation}. The RT-PCR test also suffers from a high false-positive rate during the diagnosis of COVID-19 cases \cite{ai2020correlation}, which is not acceptable. A good survey of the various image, sound, and blood test report-based datasets available for diagnosing COVID-19 cases can be found in \cite{shuja2020covid}. Recent studies \cite{zu2020coronavirus}, \cite{bernheim2020chest} have shown Chest X-Ray images of COVID-19 patients play a vital role in timely identification and further control of the COVID-19 cases. Inspired by the success of work on chest X-Ray and CT scan images, various methods and computer-aided systems have been proposed that combine deep learning methods and radiology expert knowledge to identify the COVID-19 cases. A comprehensive study of various deep learning-based methods for diagnosis of COVID-19 using chest X-Ray and CT Scan images can be found in \cite {ozturk2020automated}, \cite{khan2020coronet}, \cite{wang2020covid}, \cite{shoeibi2020automated}. The majority of the deep learning models proposed for identifying COVID-19 cases are based on transfer learning \cite{wang2020covid}, \cite{apostolopoulos2020covid}, \cite{narin2020automatic}, \cite{misra2020multi}, attention-based mechanism \cite{han2020accurate}, \cite{wang2020prior}, \cite{sharma2020covid}, \cite{mangal2020covidaid}, self-supervised learning \cite{abbas20204s}, \cite{li2020efficient} and explainable deep models \cite{karim2020deepcovidexplainer}, \cite{brunese2020explainable}, \cite{singh2020covidscreen}, \cite{wu2020jcs}. On the other hand, very little work has been performed towards the vulnerability of deep models against adversarial attacks \cite{biggio2013evasion} capable of misleading the deep model with a small perturbation in pixels of an input image. Identification of COVID-19 cases requires expert opinions over the chest X-Ray and CT scan images. It also involves the communication of the COVID-19 data through the web to receive the expert's suggestions and reports.
\par 
The deep learning models have achieved new heights of state-of-the-art (SOTA) methods in object detection \cite{erhan2014scalable}, text mining \cite{mehdiyev2017time}, speech recognition \cite{liu2017survey} and computer vision \cite{guo2016deep}. However, it has been well explored that the deep models are sensitive towards small perturbation in an input and easily fooled by the attacker. This paradigm is also known as Adversarial attack \cite{biggio2013evasion}, \cite{szegedy2013intriguing}. The study of adversarial attacks was introduced a decade ago \cite{biggio2018wild} and gained huge attention from researchers of deep learning due to the increasing demand for deep learning techniques in various real-life applications. Data and models privacy and security concerns make the study of adversarial attacks popular in deep learning research. The existence of the adversarial attacks put various questions on the generalization of deep models for the diagnosis of COVID-19 using medical images. In \cite{hirano2020vulnerability}, Hirano et al. investigated the performance analysis of deep models for the diagnosis of COVID-19 cases in the presence of adversarial attacks.  A previous study suggested the vulnerability as a major bottleneck for the medical image-based diagnosis \cite{finlayson2019adversarial}. \par 
Adversarial attacks on deep models can be subdivided into two major classes. The first type of attack is known as a white-box attack \cite{biggio2018wild}, and the second type of attack is known as a black-box attack \cite{biggio2018wild}. The white-box attacks use the full knowledge of the deep model, dataset, architecture, and parameters. However, the scenario is different in black-box attacks that only partially access the information related to deep models. The proposed method aims to provide defense against white-box attacks that are very hard to prevent in practical scenarios. Adversarial attacks are further broadly classified into two classes, targeted and untargeted attacks. The targeted attacks modify the clean images into adversarial images that make the deep model to classify the input image into a class set by the attackers. For an example, if a clean image of the non-COVID-19 case is transformed into an adversarial image with a target label set as COVID-19 case and the model classifies the images as COVID-19 instead of a normal case. On the other hand, in case of untargeted attacks, the image is transformed into an adversarial image such that the model classifies the image into labels other than the true class label of the image. For an example, the image belongs to the COVID-19 case misclassified as a normal case or pneumonia case after an untargeted adversarial attack. The work presents in this paper intents to present a defense mechanism against an untargeted class of adversarial attacks.
\par 
A pioneer work that focuses on the generation of the adversarial examples was presented by Goodfellow et al. \cite{goodfellow2014explaining}. The author proposed a fast gradient-based approach to generate adversarial samples. By taking inspiration from the initial work of Goodfellow et al. \cite{goodfellow2014explaining}, various methods have been proposed to generate adversarial examples.  Moosavi et al. \cite{moosavi2016deepfool} proposed a deep fool mechanism that generates perturbation until the confidence of the model decreases on the correct label for the given input. The iteration to create perturbation stops when the deep model is fooled. In \cite{carlini2017towards}, the author proposed an attack mechanism that uses an Adam optimization method \cite{kingma2014adam} for an adversarial attack. Sharma et al. \cite{sharma2018attend} proposed a framework that uses attention feature maps to generate adversarial examples to attack the deep model. \par 
Besides the development of adversarial attack techniques, numerous defense methodologies  \cite{miller2020adversarial}, \cite{perez2017effectiveness} have been proposed to prevent the deep models against the adversarial attacks. The defense methodologies are further grouped into two categories black box defense and white box defense. The white box defense involves adversarial images as input to train the deep model. The adversarial images are generated by one of the adversarial attack techniques \cite{goodfellow2014explaining}, \cite{moosavi2016deepfool} mentioned above. On the contrary, the black box defense does not involve the adversarial images to train the deep model that prevents adversarial attacks. Data augmentation techniques \cite{perez2017effectiveness}, input transformation \cite{guo2017countering} and an  encryption inspired shuffling of images \cite{zhao2018detecting} are among the popular techniques that are well explored to perform black-box defense. A comprehensive study of various defense techniques against the adversarial attacks can be found in \cite{miller2020adversarial}. The white box defenses are more successful as compared to the black box defense. Still, they suffer from a high probability of failure against the attacks having a complexity greater than the adversarial attacks employed to generate the adversarial images while training the white box defense models. The black box defense is independent of the complexity of the attack mechanism thus gained more attention to develop robust and secure deep models against the adversarial attacks.\par
The proposed fuzzy unique image transformation (FUIT) technique belongs to the black box defense category. To the best of our knowledge, a fuzzy logic-based black box defense has not been proposed to prevent the COVID-19 images from adversarial attacks. The two significant contributions of this paper are as follows. The first contribution is incorporating a fuzzy unique transformation method within the architecture of a deep model to secure the deep model against the adversarial attacks. The second contribution is to provide a comprehensive study of the performance of the proposed model to classify the COVID-19 cases under the various adversarial attacks. \par 
The organization of the paper is as follows: Section II presents the brief introduction of the adversarial attack and fuzzy set theory. Section III presents details of the methodology used to train the secure deep model using FUIT transformed images to prevent the adversarial attacks. Section IV presents experiments, results and ablation study, and finally, conclusions and future work are presented in Section V.
\section{Preliminaries}
This section presents a brief introduction of adversarial attack and fuzzy set.  
\subsection{\textbf{Adversarial Attacks}} The major aim of adversarial attacks \cite{biggio2013evasion} is to modify pixel values by small amount $\epsilon$. The changes that occur in a modified image are invisible for humans but well understood by deep learning models. If f denotes a function that represents a deep model with parameters $\theta$ learned using input image X and label $y$
\begin{equation}
    ~~~~~~~~~~~~~~~~~~~~~~~~~~~y=f(X,\theta)
\end{equation}
After adding small perturbation to image pixels the adversarial input image $X^{'}$ satisfies the following condition:
\begin{equation}
   ~~~~~~~~~~~~~~~~~~~~~~~~~~ ||X^{'}-X||\leq \epsilon 
\end{equation}

\begin{equation}
    ~~~~~~~~~~~~~~~~~~~~~~~~~~y'=f(X^{'},\theta)
\end{equation}
When the model is evaluated against adversarial image $X^{'}$, then $y \neq y'$, that results in a degradation in the model's performance. The phenomenon of a decrease in the classification rate of the model to classify the images is known as the adversarial attack that easily fools the model to misclassify the input image modified using a small perturbation $\epsilon$. In this paper, our primary aim is to prevent the deep model against adversarial attacks.
\subsection{\textbf{Fuzzy Set}} A set whose every element have membership value is known as fuzzy set ($\widetilde{F}$) \cite{zadeh1965fuzzy}. 
\begin{equation}
    ~~~~~~~~~~~~~~~~~~~\widetilde{F}=\left\{{(x,\mu_{\widetilde{F}}(x))~|~x\in U}\right\}
\end{equation}
Where ($\widetilde{F}$) is a fuzzy set with element x and membership value $\mu$. Here $\mu_{\widetilde{F}}(x)$ denotes membership value of x with respect to fuzzy set $\widetilde{F}$. The value of $\mu$ always lies in between $0$ to $1$. U is an universe of information.  

\section{Methodology}
This section presents details of the Fuzzy Unique Image Transformation (FUIT) technique and methodology used to build a secure deep model against the adversarial attacks.
\subsection{\textbf{Fuzzy Unique Image Transformation (FUIT)}} FUIT creates fuzzy sets from the given range of values of the image pixels. In an image where pixel values lie in a range (U) from 0 to 255. We create the R fuzzy sets that use a triangular membership function \cite{zadeh1965fuzzy} (as shown in Eq.(\ref{TRMF})) to compute the membership value ($\mu$) of the given pixel. The created fuzzy sets downsample the image pixels into an interval of range in between 1 to R. The new transformed image has pixel values between 1 to R. The FUIT technique performs discretization of the values of the image pixel into an interval $[1, R]$. 

  \begin{algorithm}[h!]
  \caption{Fuzzy Unique Image Transformation (FUIT)}\label{FUIT_Algo}
  \begin{algorithmic}[1]
    \Require Input Image ($X$), $R$-Fuzzy Sets 
    \State Initialize~$[rows, cols]$~=~size~($X$)
\For{\texttt{(i=1)\&(i<=$rows$)}}
\For{\texttt{(j=1)\&(i<=$cols$)}}
\For{\texttt{(r=1)\&(r<=$R$)}}
 \State \texttt{$Mv[r]$=$\mu_{r}(X^{ij})$}
\State Here, $Mv_{[1*R]}$= An array of membership values 
        
      \EndFor
      \State [$\mu_{max}$, $Index$]~=~$Max$ ($Mv$)
      \State $X_{F}^{ij}~=~$$Index$
      \EndFor
      \EndFor
      \State Return  $X_{F}$ (Output FUIT Transformed Image) 
  \end{algorithmic}
\end{algorithm}

\textbf{Algorithm \ref{FUIT_Algo}} shows various steps of FUIT transformation of the image $X$. 
 In this paper a triangular membership function is used to compute membership values of a pixel for the given fuzzy set. 
 \begin{equation}\label{TRMF}
     ~~~~~~~~~\mu (x,p,q,r)=\begin{cases}0 & x\leq p\\\frac{x-p}{q-p} &  p \leq x\leq q \\\frac{r-x}{r-q}&q \leq x\leq r \\0& r \leq x\end{cases}
 \end{equation}
 Eq.(\ref{TRMF}) shows the triangular membership function with three parameters $p, q, r$ and input $x$. $\mu$ denotes triangular membership value.  
  
%\begin{algorithm}
%\\
%\KwIn{Input:~Input Image ($X$), Number of Fuzzy Sets ($R$)}\\\\
%\KwOut{Output:~Transformed Image ($X_{F}$})\\
%\State Initialize~$[rows, cols]$~=~size~($X$)\\
%\For{\texttt{(i=1)\&(i<=$rows$)}}\\
%\For{\texttt{(j=1)\&(i<=$cols$)}}\\
%\For{\texttt{(r=1)\&(r<=$R$)}}\\
% \State \texttt{$Mv[r]$=$\mu_{r}(X^{ij})$}\\
%\State Here, $Mv_{[1*R]}$= array of membership values \\
        
%      \EndFor\\
%      \State [$\mu_{max}$, $Index$]~=~maximum ($Mv$)\\
%      \State $X_{F}^{ij}~=~$$Index$\\
%      \EndFor\\
%      \EndFor\\
%      \State Return  $X_{F}$\\
%\caption{Fuzzy Unique Image Transformation (FUIT)}
%\label{FUIT_Algo}
%\end{algorithm}

%%%%%%%%%%%%%%%%%%%%%%%%%%%%%%%%%%%

%%%%%%%%%%%%%%%%%%%%%%%%%%%%%%%%%%%%%%

\begin{figure}[h!]
	\includegraphics[width=9 cm,height=7 cm]{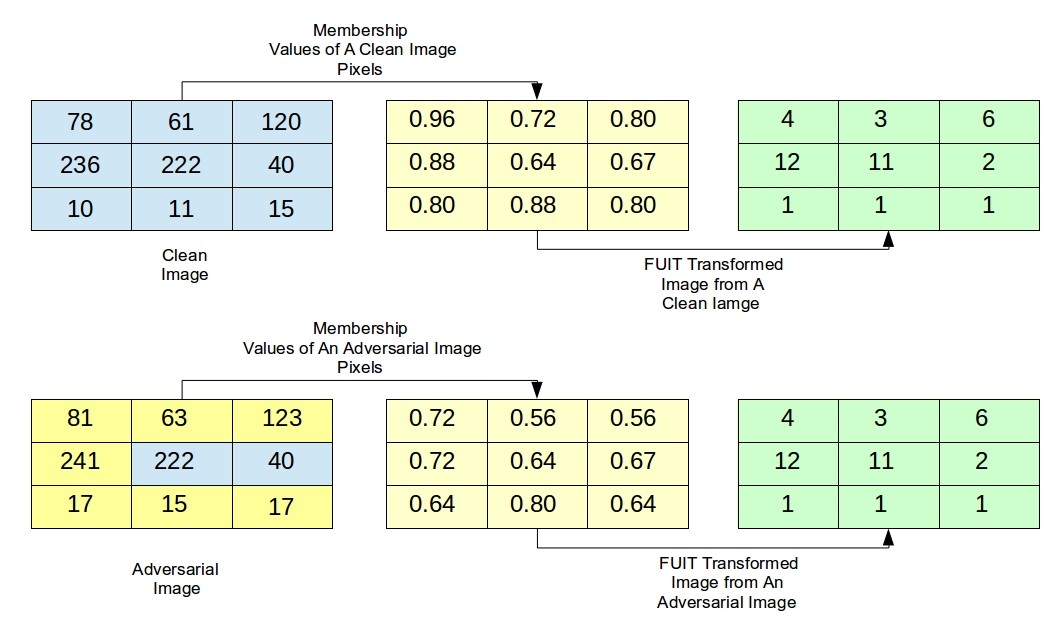}
	\caption{Various steps of Fuzzy Unique Image Transformation}
	\label{FUIT}
	\end{figure}

\begin{figure*}[h!]
 	\includegraphics[width=17 cm,height=6 cm]{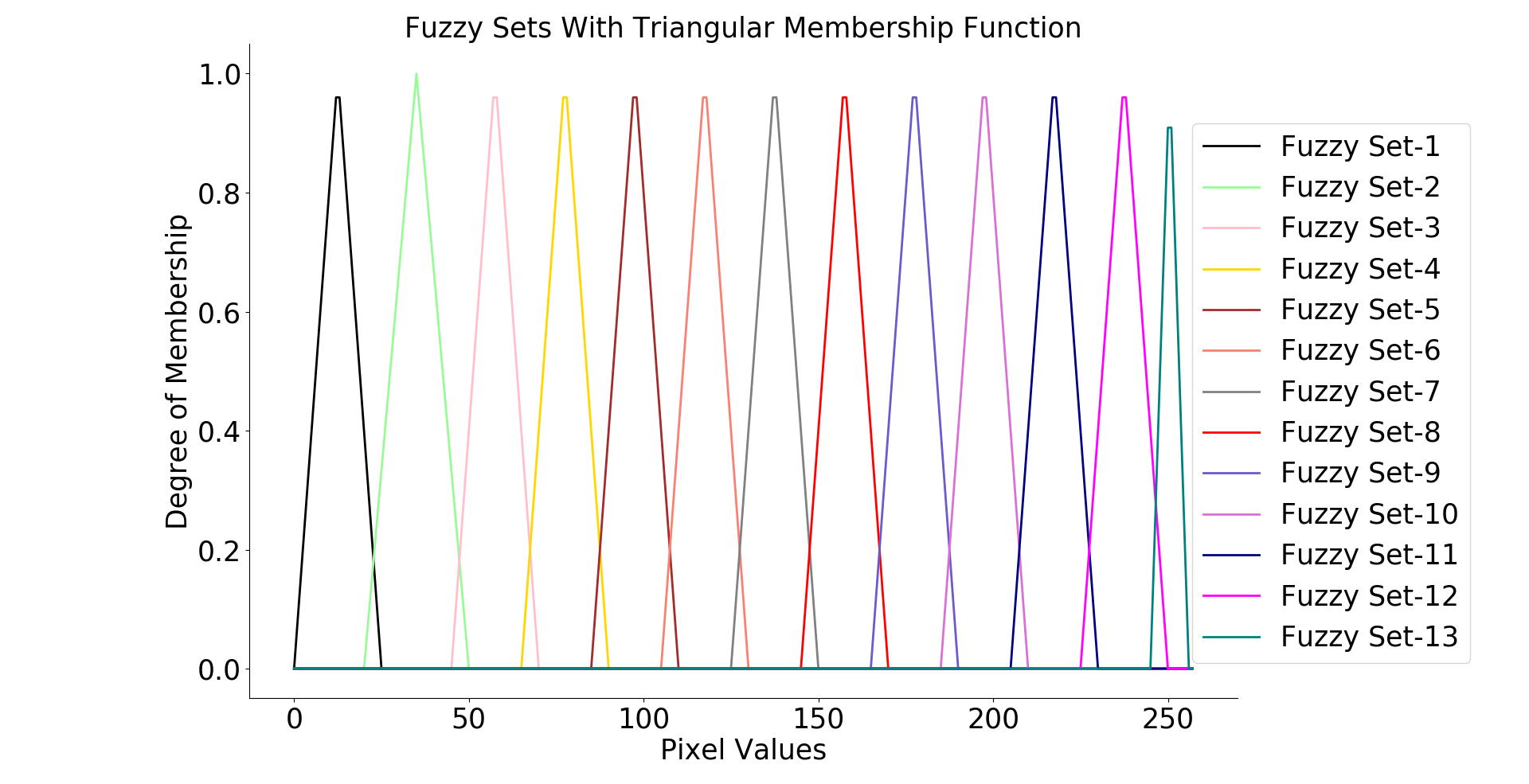}
	\caption{Fuzzy Sets Created for FUIT}
	\label{Fuzzy_Sets}
	\end{figure*}

For an example, consider an image of size (3*3) having values $[78,61,120,236,222,40,10,11,15] $ as shown in Fig.(\ref{FUIT}). The image is transformed into FUIT image using the \textbf{Algorithm \ref{FUIT_Algo}}.	
The new image created after the FUIT technique has pixel values $[4,3,6,12,11,2,1,1,1]$ based on membership values $[0.96,0.72,0.80,0.88,0.64,0.67,0.80,0.88,0.80]$ computed from the created fuzzy sets. Fig.(\ref{Fuzzy_Sets}) shows various fuzzy sets created for the FUIT transformation. After the previous image transformed into an adversarial image the new pixel values become  $[81,63,123,241,222,40,17,15,17]$.  After applying the FUIT algorithm the new transformed image becomes $[4,3,6,12,11,2,1,1,1]$  with  membership values $[0.72,0.56,0.56,0.72,0.64,0.67,0.64,0.80,0.64]$. \par 

\begin{figure*}[h!]
	\includegraphics[width=15 cm,height=6cm]{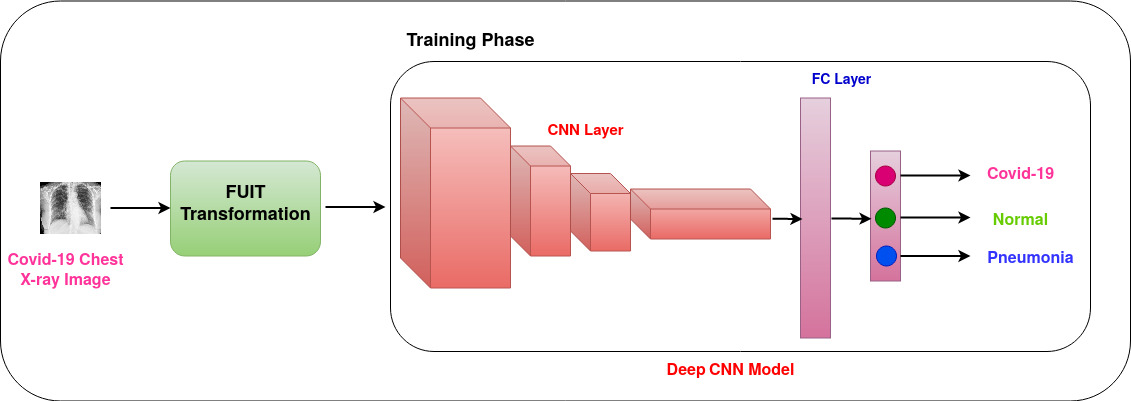}
	\centering
	\caption{ Training of Deep Model with COVID-19 Chest X-ray Images Transformed by FUIT Technique}
	\label{Train_Flow}
	\end{figure*}

\begin{figure*}[h!]
	\includegraphics[width=15 cm,height=6 cm]{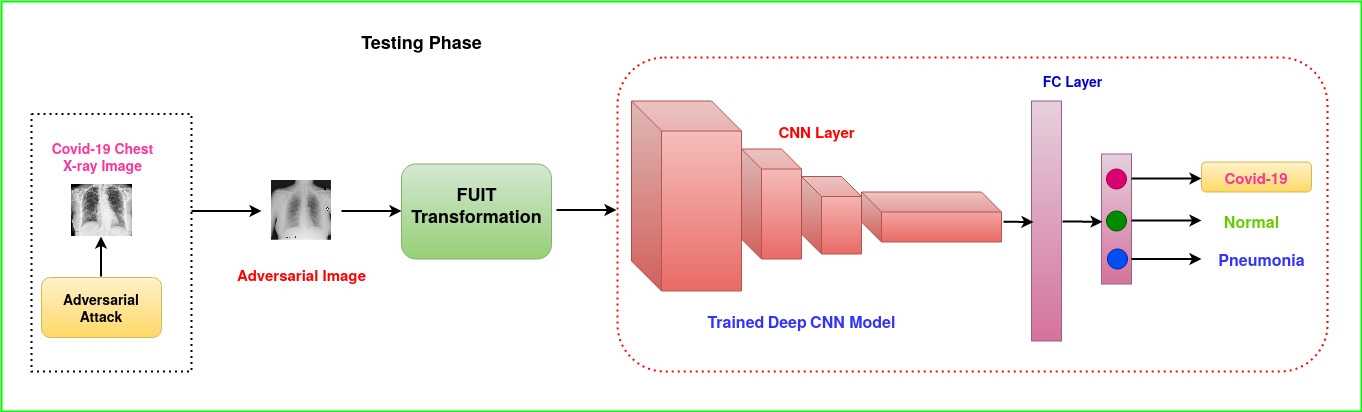}
	\centering
	\caption{Classification of Adversarial COVID-19 Chest X-ray Image Transformed by FUIT Technique}
	\label{Test_Flow}
	\end{figure*}

It is clear from the  Fig.(\ref{FUIT}) that the image persists its own characteristics in both the situations (for a clean image or under attack). The characteristics of the images can be expressed in terms of the number of unique pixel values $(V)$. For a clean image $V=[4,3,6,12,11,2,1]$ and for an adversarial image $V=[4,3,6,12,11,2,1]$. In both the cases $||V||=7$. The FUIT transformation prevents the increase in the value of $||V||$ due to adversarial attack. In other words, a variance of $||V||$ remains the same for the clean image and image under adversarial attack. The increase in the variance of $||V||$ due to adversarial attack easily fools the deep model and results in a high misclassification rate. The FUIT transformation is deployed before forwarding the image as input to the deep model. The model easily learns the FUIT transformed images and shows secure nature towards the adversarial attacks. All the training images are prepossessed with FUIT technique while training of the deep model and all test images i.e., clear images or adversarial images, are also pre-processed with FUIT technique before classification by the deep model. Fig.(\ref{Train_Flow}) and Fig.(\ref{Test_Flow}) show the overall flow of classification and defense mechanism used against the adversarial attacks on deep COVID-19 model using the proposed framework. 
	
%%%%%%%%%%%%%%%%%%%%%%%%%%%%%%%%%%
\section{Experiments and Results}
This section presents a description of datasets and the details of the model and results compared with the previous proposal for COVID-19.
\subsection{\textbf{COVID-19 Chest X-Ray Image Dataset}} 
\begin{figure*}[h!] 
	\centering
	\begin{subfigure}[h]{.4\textwidth}
		\includegraphics[width=6.5 cm,height=4 cm]{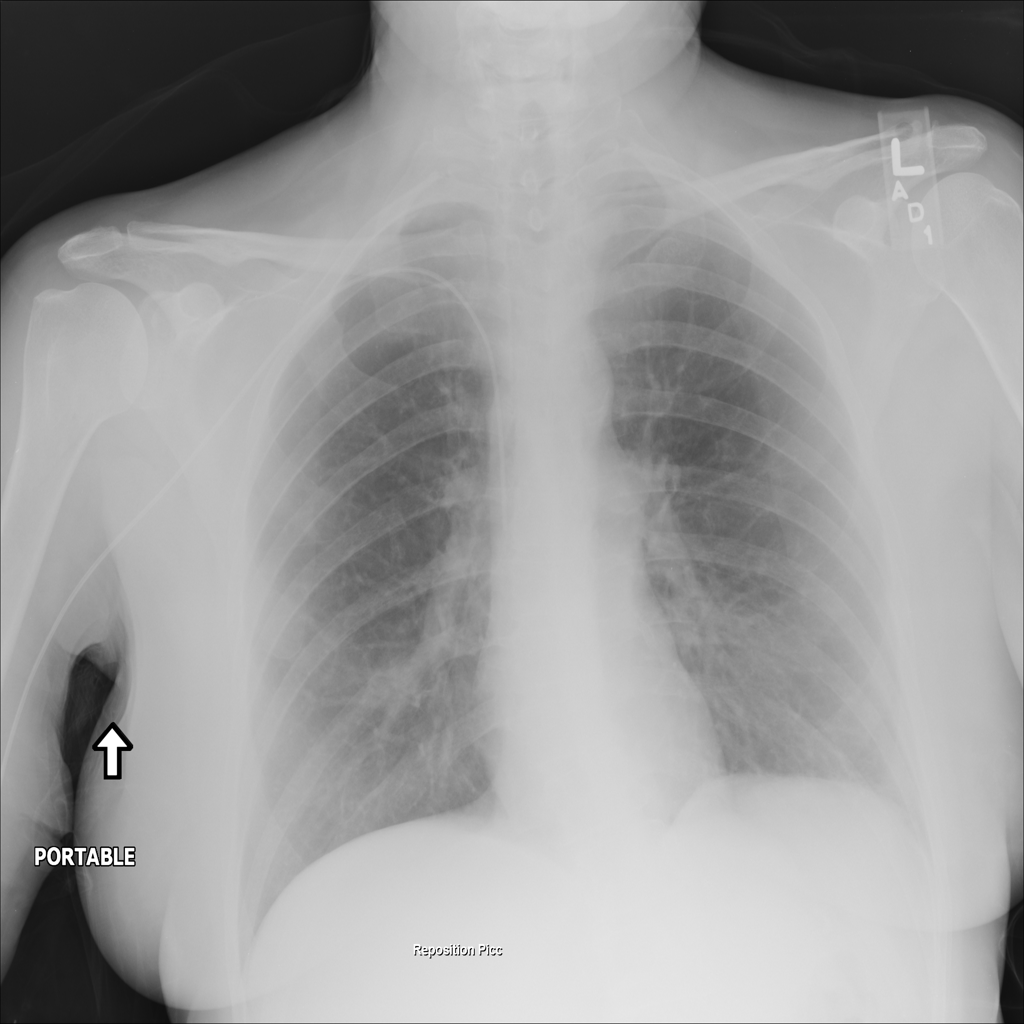}
		\caption{Normal Patient-1 }
		\label{N1}
	\end{subfigure} %\qquad
	\begin{subfigure}[h]{.4\textwidth}
		\includegraphics[width=6.5 cm,height=4 cm]{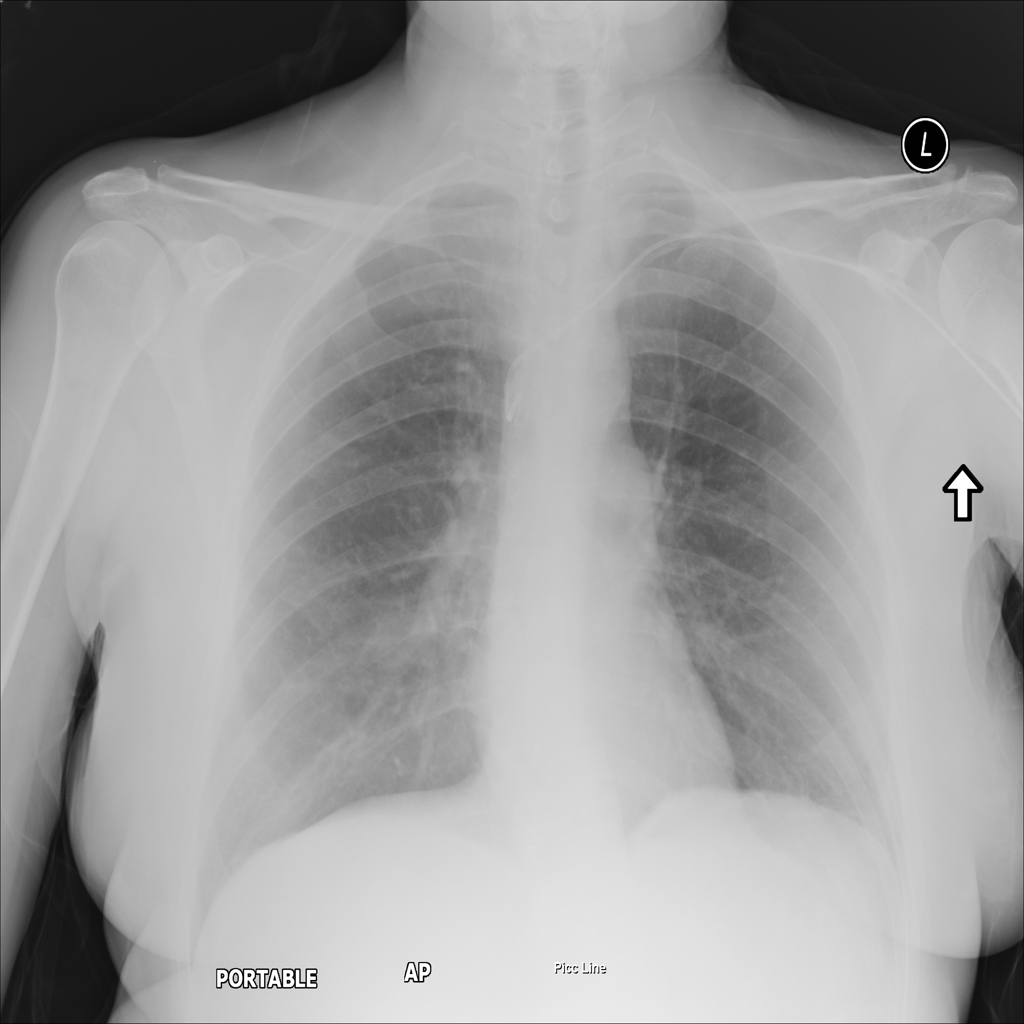}
		\caption{Normal Patient-2}
		\label{N2}
	\end{subfigure} %\qquad
	\begin{subfigure}[h]{.4\textwidth}
		\includegraphics[width=6.5 cm,height=4 cm]{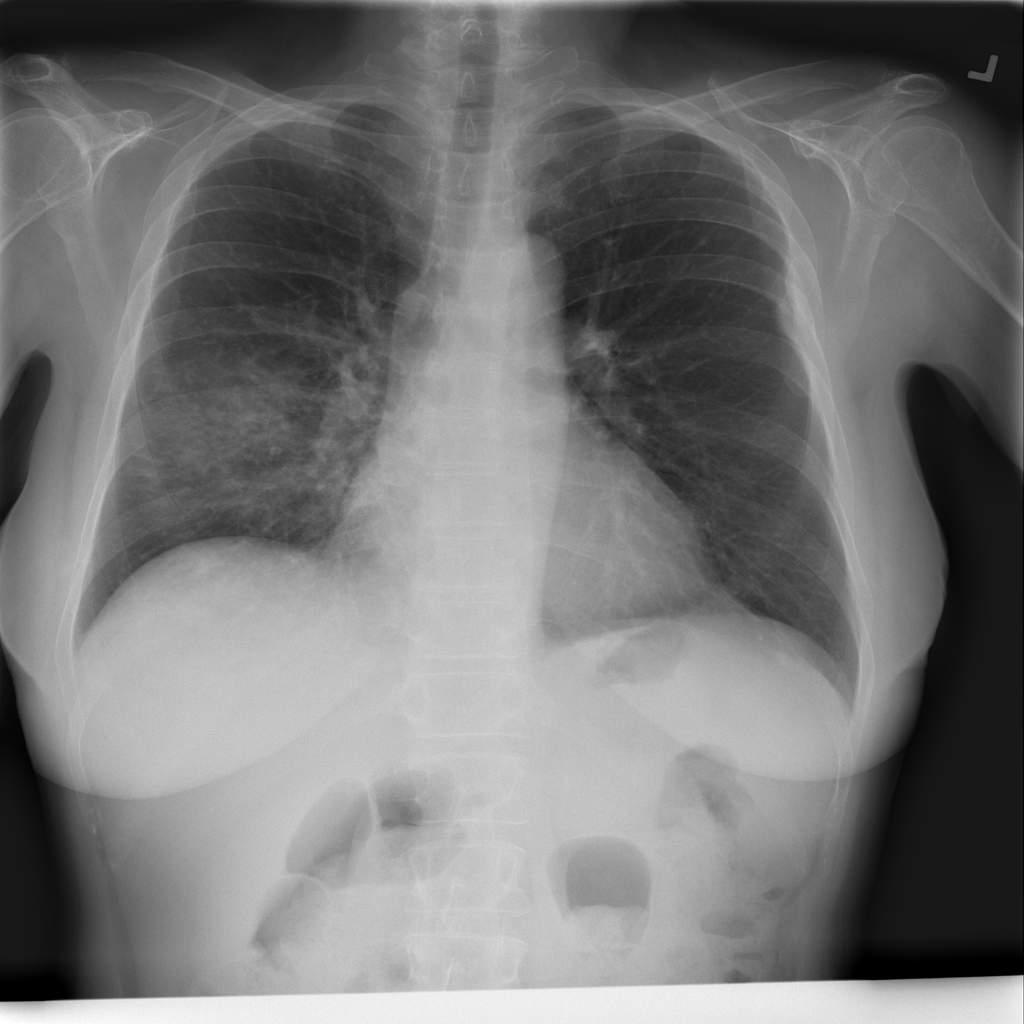}
		\caption{Pneumonia Patient-1 }
		\label{PN1}
	\end{subfigure} %\qquad
	\begin{subfigure}[h]{.4\textwidth}
		\includegraphics[width=6.5 cm,height=4 cm]{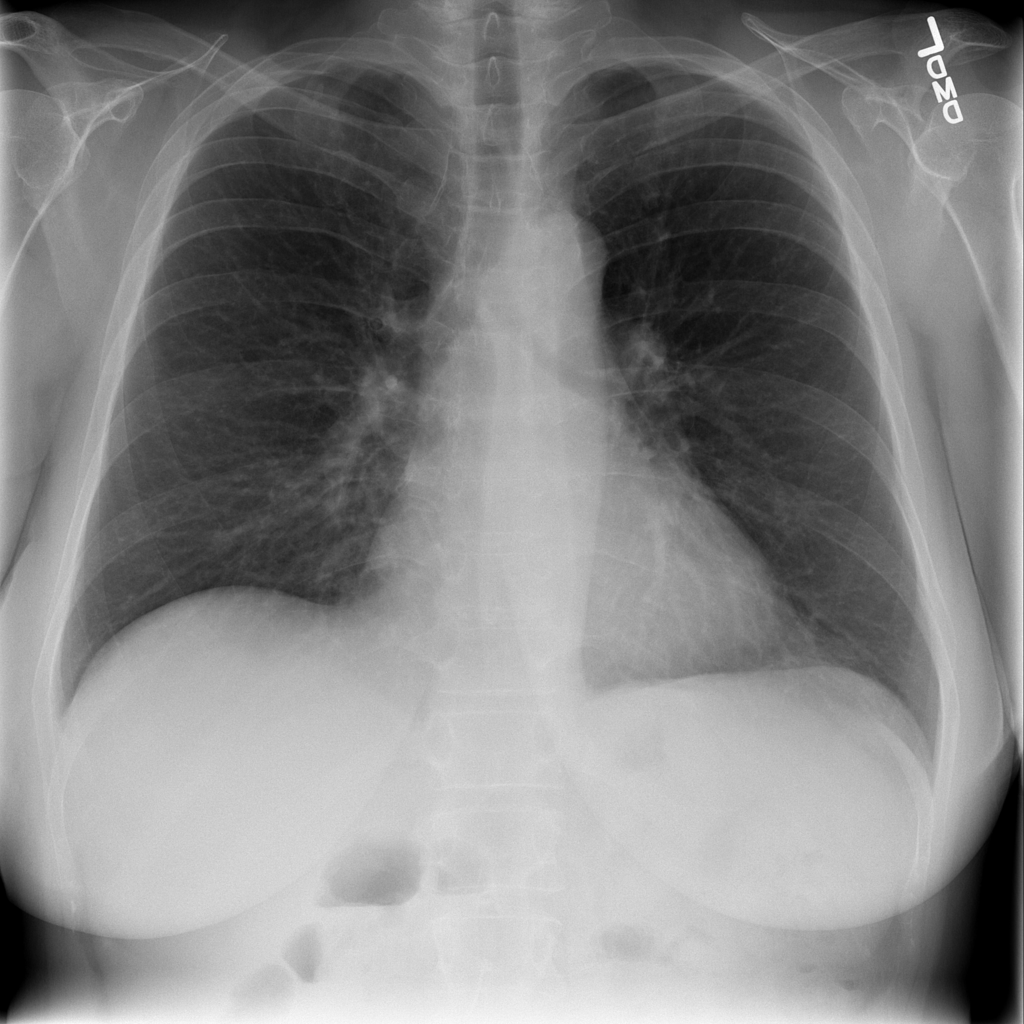}
		\caption{Pneumonia Patient-2}
		\label{PN2}
	\end{subfigure}
	
	\begin{subfigure}[h]{.4\textwidth}
		\includegraphics[width=6.5 cm,height=4 cm]{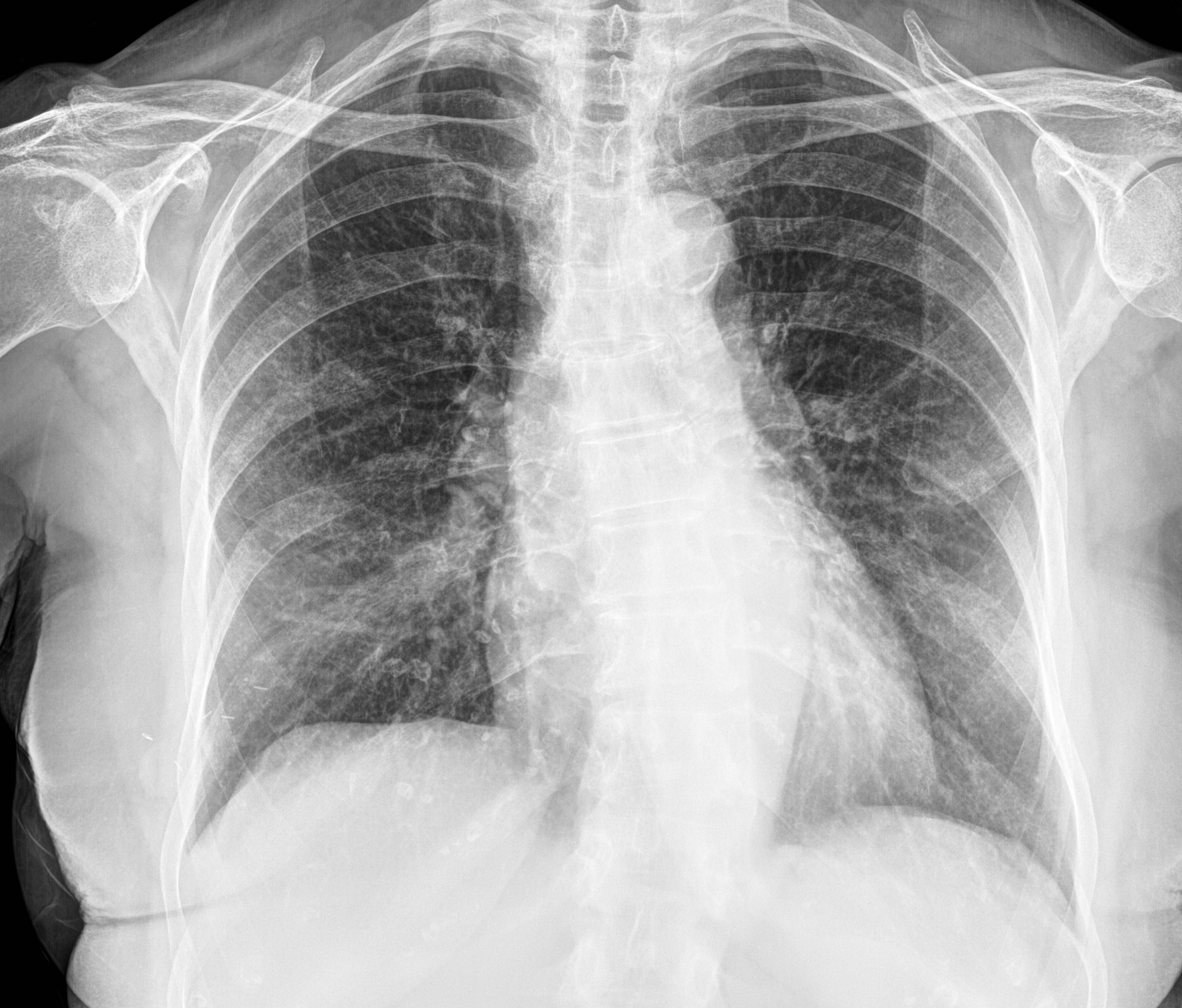}
		\caption{COVID-19 Patient-1}
		\label{Covid-1}
	\end{subfigure}
	\begin{subfigure}[h]{.4\textwidth}
		\includegraphics[width=6.5 cm,height=4 cm]{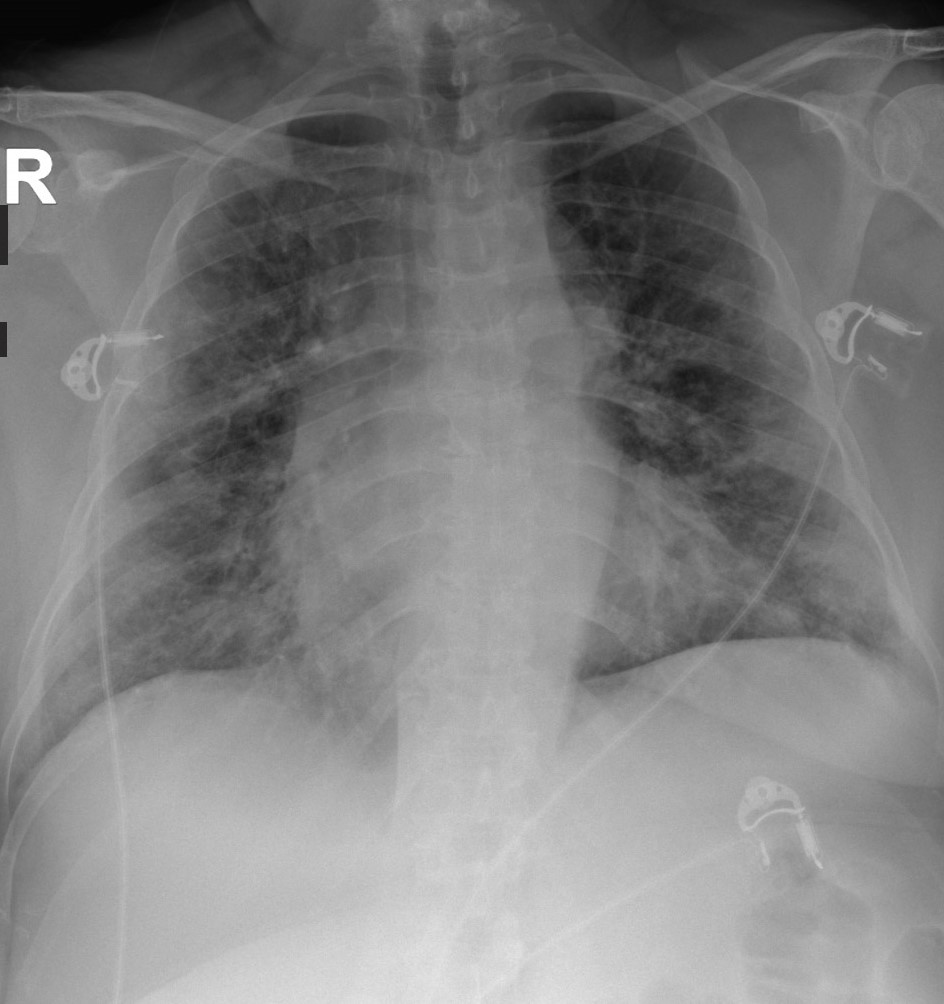}
		\caption{COVID-19 Patient-2}
		\label{Covid-2}
	\end{subfigure} %\qquad
	\caption{Sample Images of Normal, COVID-19 and Pneumonia Cases From Chest X-Ray Image Dataset \cite{cohen2020covid}}\label{Dataset}
\end{figure*}

The performance of the proposed method is first evaluated on COVID-19 Chest X-Ray Image Dataset \cite{cohen2020covid}. The data set includes a collection of chest X-Ray images of people belonging to Normal, Pneumonia, and COVID-19 classes. Several contributions from people belonging to different places increase the size of the dataset. At the time of this study, the dataset contains a total of 1125 images. Among the available 1125 images, 500 images belong to Normal Class, 500 images belong to people suffering from pneumonia, and the remaining 125 images belong to people infected from COVID-19. The study followed 5-fold cross-validation to evaluate the performance of the proposed framework. Fig. (\ref{Dataset}) shows sample images of the chest X-Ray of persons belong to normal, pneumonia, and COVID-19 classes.

\subsection{\textbf{Model Description}} Total of nine models (M1-M9) are trained using weights initialized with three different pre-trained models i.e.  Resnet-18 \cite{he2016deep}, VGG-16 \cite{simonyan2014very} and GoogLeNet \cite{szegedy2015going} respectively. Table \ref{Chest_Models} shows details of the nine models prepared for comparison.
\begin{table}[h!]\setlength{\extrarowheight}{3pt}\caption{Description of Different Models Developed for Chest X-Ray Dataset}\label{Chest_Models}
\scalebox{0.85}{
\begin{tabular}{|c|c|c|c|}
\hline
\textbf{Model} & \textbf{Dataset} & \textbf{Image Type}   & \textbf{Pre-Trained Model} \\ \hline
M1             & Chest X-Ray      & Clean Image                    & Resnet-18                  \\ \hline
M2             & Chest X-Ray      & Clean Image                    & VGG-16                     \\ \hline
M3             & Chest X-Ray      & Clean Image                    & GoogLeNet                  \\ \hline
M4             & Chest X-Ray      & FUIT Transformed Image         & Resnet-18                  \\ \hline
M5             & Chest X-Ray      & FUIT Transformed Image         & VGG-16                     \\ \hline
M6             & Chest X-Ray      & FUIT Transformed Image         & GoogLeNet                  \\ \hline
M7             & Chest X-Ray      & Discretization Transformed Image & Resnet-18                  \\ \hline
M8             & Chest X-Ray      & Discretization Transformed Image & VGG-16                     \\ \hline
M9             & Chest X-Ray      & Discretization Transformed Image & GoogLeNet                  \\ \hline
\end{tabular}}
\end{table}
The models (M1-M9) are further evaluated against the adversarial images created using six different types of attacks exist in literature. Table \ref{AttacksUsed} shows the six different attacks that are used in this study. Parameters of all the six attacks mentioned in the Table \ref{AttacksUsed} are shown in Table \ref{Attack_Para}.
\begin{table}[h!]\setlength{\extrarowheight}{3pt}\caption{Details of Different Adversarial Attacks}\label{AttacksUsed}
\begin{tabular}{|c|c|}
\hline
\textbf{S.No.} & \textbf{Attacks}                                                                               \\ \hline
1              & Deep Fool \cite{moosavi2016deepfool}                                          \\ \hline
2              & Fast Gradient Sign Attack \textbf{(FGSM)} \cite{goodfellow2014explaining}              \\ \hline
3              & Basic Iterative Method \textbf{(BIM)} \cite{kurakin2016adversarial}                    \\ \hline
4              & Carlini \& Wagner \textbf{(CW)} \cite{carlini2017towards}                              \\ \hline
5              & Projected Gradient Descent With Random Start \textbf{(PGD-r)}  \cite{madry2017towards}  \\ \hline
6              & Projected Gradient Descent Without Random Start   \textbf{(PGD)} \cite{madry2017towards} \\ \hline
\end{tabular}
\end{table}
\begin{table}[h!]\setlength{\extrarowheight}{3pt}\caption{Parameters of Six Adversarial  Attacks}\label{Attack_Para}
~~~~~~~\begin{tabular}{|c|c|}
\hline
\textbf{Attack}    & \textbf{Parameters}                                   \\ \hline
\textbf{PGD}       & $\epsilon$=0.3, $\alpha$=4/255, Steps=40                    \\ \hline
\textbf{PGD-r}     & $\epsilon$=0.3, $\alpha$=4/255, Steps=40, Random Start=True \\ \hline
\textbf{FGSM}      & $\epsilon$=0.008                                         \\ \hline
\textbf{CW}        & C=2, Kappa=2, Steps=500, learning rate=0.01           \\ \hline
\textbf{Deep Fool} & Steps=20                                              \\ \hline
\textbf{BIM}       & $\epsilon$=8/255, $\alpha$=1/255, steps=10                  \\ \hline
\end{tabular}
\end{table}
\subsection{\textbf{Experimental Settings}}
The early stop technique is used to train the models, and the maximum epoch is set as 150. The learning rate is 0.001, and the batch size is selected as 32. All experiments are conducted on Ubuntu 16.04 LTS operating system with 16 GB RAM and NVIDIA GM107M 4 GB GPU. All scripts are developed using an open-source Pytorch 1.4 library. 
All images are resized to a size required by the three pre-trained models used to initialize the deep model weights. The deep model is trained using an Adam optimization \cite{kingma2014adam}. 
\subsection{\textbf{Loss Function}}
A loss function $\mathcal{L}$ is selected as cross entropy loss as shown in Eq.(\ref{Loss}). Here x is an input and $C$ is class label. $k$ is the total number of classes.
\begin{equation}\label{Loss}
    ~~~~~~~~~~\mathcal{L}(x,C)=-log\left(\frac{exp~(x[C])}{\sum_kexp(x[k])}\right)
\end{equation}
\subsection{\textbf{Results on Chest X-Ray Dataset}} 
As mentioned earlier for the comparative analysis total of nine models are developed. We evaluated the performance of the proposed model in two settings. Models are trained for binary and three class classification scenarios. In the case of the binary classification, images belong to COVID-19 and pneumonia classes are considered to come from the same class. Initially, M1, M2, and M3 are trained and tested on the clean chest X-Ray images. Table \ref{Accuracy_Clean_Three} shows the performance of the three models for binary and three class classification. For the binary classification, the model M1 yields the highest mean accuracy of 97.28\%, and the model M3 shows the lowest mean accuracy of 96.84\%. The model M2 achieves the accuracy of 97.14\%. For the three-class classification, the model M1 shows the highest mean accuracy of 88.12\%. The model M3 shows the lowest mean accuracy of 87.03\%. \par 
\begin{table}[h!]\setlength{\extrarowheight}{3pt}\caption{Classification Accuracy of Different Models ON COVID-19 Chest X-ray Dataset}\label{Accuracy_Clean_Three}
~~~~~~\begin{tabular}{|c|c|c|c|}
\hline
\multirow{2}{*}{\textbf{Model}}                                 & \multicolumn{3}{c|}{\textbf{Accuracy (\%)}}                   \\ \cline{2-4} 
                                                                & \textbf{M1} & \textbf{M2} & \textbf{M3} \\ \hline
\textbf{\begin{tabular}[c]{@{}c@{}}Binary  Class\end{tabular}} & 97.28  $\pm{0.21} $           & 97.14  $\pm{0.30} $        & 96.84 $\pm{0.33}$           \\ \hline
\textbf{\begin{tabular}[c]{@{}c@{}}Three Class\end{tabular}}  & 88.12 $\pm{0.27}  $           & 87.81  $\pm{0.36}$        & 87.03  $\pm{0.40}$           \\ \hline
\end{tabular}
\end{table}
\begin{table}[h!]\setlength{\extrarowheight}{3pt}\caption{Adversarial Attack on Models Trained Using Clean COVID-19 Chest X-Ray Images}\label{Attacks_Clean}
~~~\begin{tabular}{|c|c|c|c|c|c|c|}
\hline
\multirow{2}{*}{\textbf{Attack}} & \multicolumn{3}{c|}{\textbf{Accuracy (Binary Class)}}     & \multicolumn{3}{c|}{\textbf{Accuracy (Three Class)}}     \\ \cline{2-7} 
                                 & \textbf{M1} & \textbf{M2} & \textbf{M3} & \textbf{M1} & \textbf{M2} & \textbf{M3} \\ \hline
\textbf{PGD}                     & 54.34              & 51.17           & 50.46              & 47.12              & 46.74           & 44.58              \\ \hline
\textbf{PGD-r}                   & 50.12              & 48.31           & 48.12              & 40.17              & 39.51           & 39.23              \\ \hline
\textbf{FGSM}                    & 47.12              & 46.18           & 46.05              & 39.14              & 38.74           & 37.12              \\ \hline
\textbf{CW}                      & 10.74              & 10.25           & 10.05              & 9.81               & 9.19            & 8.89               \\ \hline
\textbf{Deep Fool}               & 14.61              & 13.92           & 13.84              & 12.29              & 11.81           & 11.01              \\ \hline
\textbf{BIM}                     & 9.17               & 8.69            & 8.61               & 8.15               & 7.10            & 7.01               \\ \hline
\end{tabular}
\end{table}
After evaluation of the performance of the models M1,M2 and M3 on the clean Chest X-Ray images, the models are tested against the adversarial images generated from six different attacks, as listed in the Table \ref{AttacksUsed}. Table \ref{Attacks_Clean} shows the performance of the three models for binary and three class classification. The model M1 shows an accuracy of 54.34\% in the presence of a PGD attack. The model M1 shows the accuracy of 50.12\%, 47.12\%, 10.74\%, 14.61\%, and 9.17\% for PGD-r, FGSM, CW, Deep Fool, and BIM attacks, respectively. The BIM attack is the most successful attack that results in the lowest accuracy of 9.17\%.  For the same attack, M2 and M3 show an accuracy of 8.69\% and 8.61\%, respectively. For three-class classification, the lowest accuracy of M1, M2, and M3 are 8.15\%, 7.10\%, and 7.01\%, respectively. It is clear from the Table  \ref{Attacks_Clean} that the models trained on clean images perform poorly and highly insecure against the adversarial images generated by the six adversarial attacks. 
\par 
 Again three models M4, M5 and M6 are trained for binary and three class-classifications using the images obtained after the FUIT transformation. Table \ref{Accuracy_FUIT_Three} shows the accuracy of models M4, M5, and M6 when trained and test on FUIT transformed images. The accuracy of model M4 is 96.81\% and 87.25\% for binary and three class classification and little lower than the model M1. It is clear from Table \ref{Accuracy_FUIT_Three} that the FUIT transformed images are learnable by the deep model. 
\begin{table}[h!]\setlength{\extrarowheight}{3pt}\caption{Classification Accuracy of Different Models ON FUIT Transformed COVID-19 Chest X-ray Dataset}\label{Accuracy_FUIT_Three}
~~~~\begin{tabular}{|c|c|c|c|}
\hline
\multirow{2}{*}{\textbf{Model (FUIT)}}                                 & \multicolumn{3}{c|}{\textbf{Accuracy (\%)}}                                   \\ \cline{2-4} 
                                                                & \textbf{M4} & \textbf{M5} & \textbf{M6} \\ \hline
\textbf{\begin{tabular}[c]{@{}c@{}}Binary Class\end{tabular}} & 96.81  $\pm{0.14} $                 & 96.27  $\pm{0.26} $              & 96.03   $\pm{0.31} $                \\ \hline
\textbf{\begin{tabular}[c]{@{}c@{}}Three Class\end{tabular}}  & 87.25   $\pm{0.17} $                & 86.95   $\pm{0.25} $             & 86.12   $\pm{0.29} $                \\ \hline
\end{tabular}
\end{table}

\begin{table}[h!]\setlength{\extrarowheight}{3pt}\caption{Adversarial Attack on Models Trained Using FUIT Transformed COVID-19 Chest X-Ray Images}\label{Attacks_FUIT}
~~~\begin{tabular}{|c|c|c|c|c|c|c|}
\hline
\multirow{2}{*}{\textbf{Attack}} & \multicolumn{3}{c|}{\textbf{Accuracy (Binary Class)}}     & \multicolumn{3}{c|}{\textbf{Accuracy (Three Class)}}     \\ \cline{2-7} 
                                 & \textbf{M4} & \textbf{M5} & \textbf{M6} & \textbf{M4} & \textbf{M5} & \textbf{M6} \\ \hline
\textbf{PGD}                     & 96.54              & 96.20           & 95.97              & 87.23              & 86.81           & 85.49              \\ \hline
\textbf{PGD-r}                   & 96.47              & 96.13           & 95.59              & 87.12              & 86.79           & 85.41              \\ \hline
\textbf{FGSM}                    & 96.49              & 96.17           & 95.42              & 87.09              & 86.70           & 85.27              \\ \hline
\textbf{CW}                      & 96.12              & 95.89           & 95.14              & 86.73              & 86.67           & 85.26              \\ \hline
\textbf{Deep Fool}               & 95.91              & 95.49           & 95.06              & 86.62              & 86.60           & 85.19              \\ \hline
\textbf{BIM}                     & 95.31              & 95.28           & 95.01              & 86.59              & 86.47           & 85.16              \\ \hline
\end{tabular}
\end{table}

% \usepackage{multirow}

% \usepackage{multirow}

% \usepackage{multirow}

\begin{figure}[!h]
	~~~\includegraphics[width=8 cm,height=5cm]{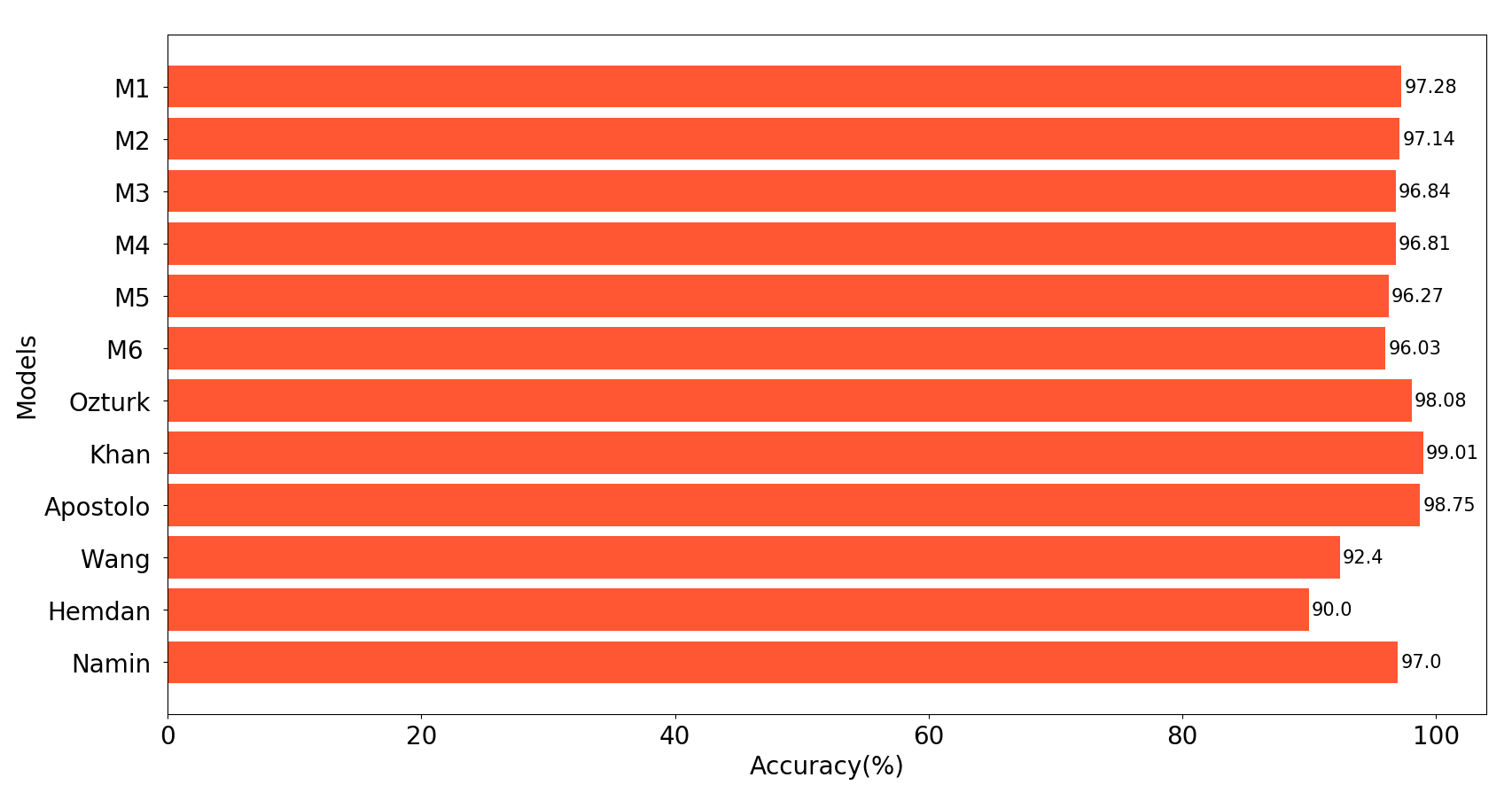}
	\caption{Comparison of Accuracy of Different Models for Binary Class Classification to Classify Chest X-Ray Dataset}
	\label{Binary_Chest}
	\end{figure}

\begin{figure}[!h]
	~~~\includegraphics[width=8 cm,height=5cm]{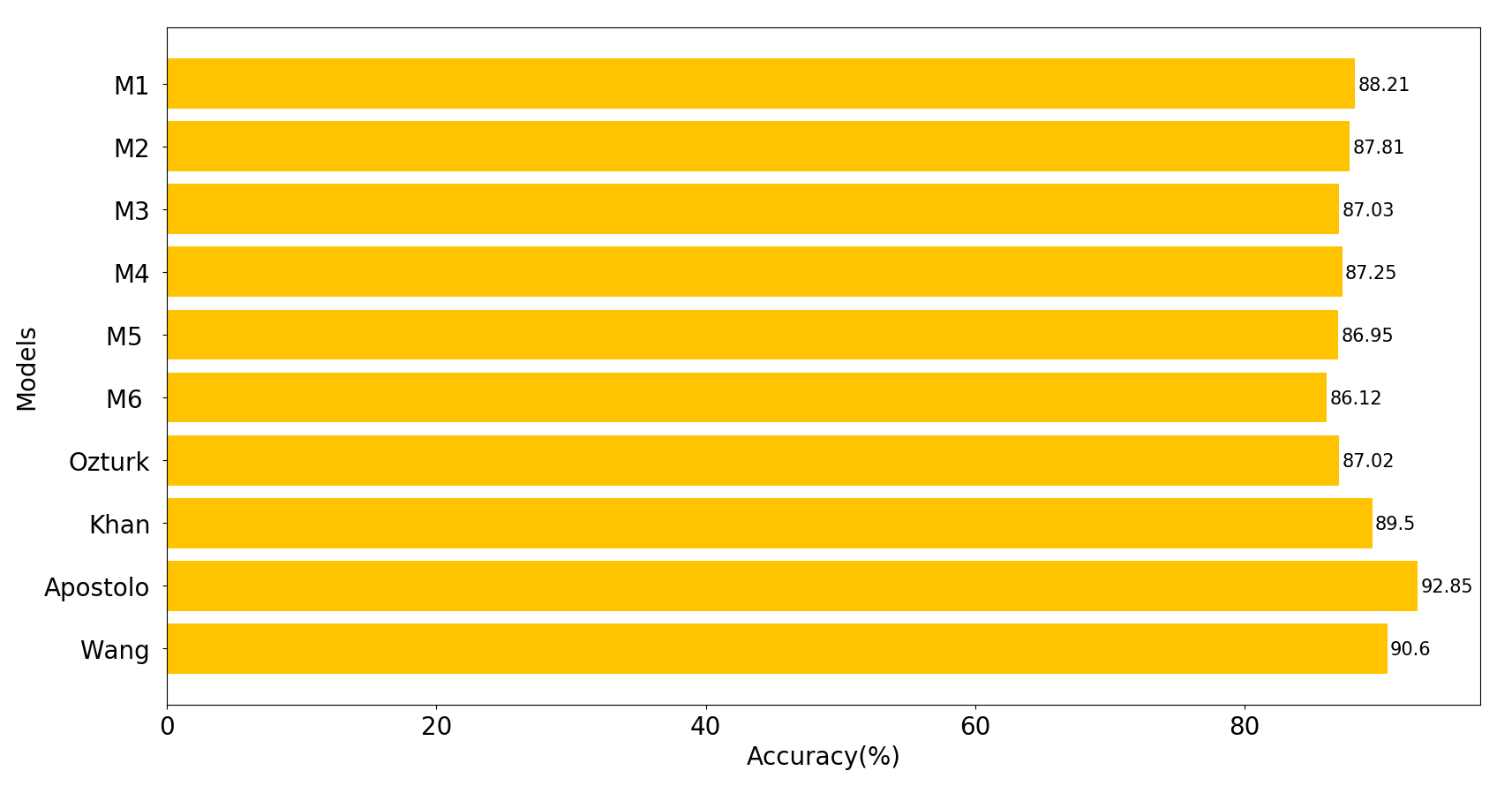}
	\caption{Comparison of Accuracy of Different Models for Three Class Classification to Classify Chest X-Ray Dataset}
	\label{Three_Chest}
	\end{figure}

The models M4, M5, and M6 are tested against the adversarial images generated using the six attacks. It is worth noting here all the training, test, or adversarial images are first transformed into FUIT technique and then given as input to the deep model. Table \ref{Attacks_FUIT} shows the accuracy achieved by the three models in the presence of six adversarial attacks. The model M4 shows the highest accuracy of 96.54\% and 87.23\% for the binary and three class classification when tested for the PDG attack. The same model also shows an accuracy of 95.13\% and 86.59\% against the most successful attack i.e., BIM attack. It is clear from Table \ref{Attacks_FUIT} that the proposed FUIT technique prevents the deep model against the adversarial attacks and and persist high accuracy to classify the COVID-19 cases when attacked with the adversarial images.. Table \ref{SOTA_Binary} and Table \ref{SOTA_Three} show a comparison of the developed models with the state-of-the-art (SOTA) methods for the diagnosis of COVDI-19 cases using the chest X-Ray images. The accuracy of the proposed FUIT transformation-based model is comparable to SOTA models and provides reliable security against adversarial attacks. Fig.(\ref{Binary_Chest}) and Fig.(\ref{Three_Chest}) show comparison of different models to detect the COVID-19 cases in binary and three class classification scenarios respectively. 
\begin{table}[h!]\setlength{\extrarowheight}{3pt}\caption{Comparison of Accuracy of SOTA Methods for Binary Classification }\label{SOTA_Binary}
	~~~~~~~~~~~~~~~~~\begin{tabular}{|c|c|}
		\hline
		\textbf{Model} & \textbf{\begin{tabular}[c]{@{}c@{}}Mean (\%)\end{tabular}} \\ \hline
		M1   & 97.28                                                                 \\ \hline
		M2    & 97.14                                                       \\ \hline
		M3             & 96.84                                                                 \\ \hline
		M4             & 96.81                                                                 \\ \hline
		M5             & 96.27                                                                  \\ \hline
		M6             & 96.03                                                                  \\ \hline
		Ozturk et al. \cite{ozturk2020automated}  & 98.08                                                                  \\ \hline
		Khan et al. \cite{khan2020coronet}   & 99.01                                                                 \\ \hline
		Apostolopoulos et al. \cite{apostolopoulos2020covid}   & 98.75                                                                 \\ \hline
		Wang et al. \cite{wang2020covid}   & 92.40                                                                \\ \hline
		Hemdan et al. \cite{hemdan2020covidx}   & 90                                                               \\ \hline
		Narnin et al. \cite{narin2020automatic}   & 97                                                               \\ \hline
	\end{tabular}
\end{table}

\begin{table}[h!]\setlength{\extrarowheight}{3pt}\caption{Comparison of Accuracy of SOTA Methods  for Three Class Classification }\label{SOTA_Three}
	~~~~~~~~~~~~~~~~~~~~\begin{tabular}{|c|c|}
		\hline
		\textbf{Model} & \textbf{\begin{tabular}[c]{@{}c@{}}Mean  (\%)\end{tabular}} \\ \hline
		M1    & 88.21                                                                  \\ \hline
		M2   & 87.81                                                         \\ \hline
		M3             & 87.03                                                                  \\ \hline
		M4             & 87.25                                                                  \\ \hline
		M5             & 86.95                                                                  \\ \hline
		M6             & 86.12                                                                  \\ \hline
		Ozturk et al. \cite{ozturk2020automated}  & 87.02                                                                  \\ \hline
		Khan et al. \cite{khan2020coronet}   & 89.50                                                                 \\ \hline
		Apostolopoulos et al. \cite{apostolopoulos2020covid}   & 92.85                                                                 \\ \hline
		Wang et al. \cite{wang2020covid}   & 90.60                                                               \\ \hline
		
	\end{tabular}
\end{table}


\subsection{\textbf{Ablation Study and Discussion}}
Apart from evaluating the performance of a deep model on the FUIT processed images, the study of the performance of the deep model trained on images transformed using the typical discretization is also evaluated. Table \ref{Accuracy_DT_Three} shows the accuracy of models M7, M8, and M9 trained for binary and three class classification using the images transformed with the typical discretization.  In typical discretization, the range of value of pixels in images is divided into intervals.
\begin{table}[h!]\setlength{\extrarowheight}{3pt}\caption{Classification Accuracy of Different Models On discretization-based Transformed COVID-19 Chest X-ray Dataset}\label{Accuracy_DT_Three}
\begin{tabular}{|c|c|c|c|}
\hline
\multirow{2}{*}{\textbf{Model (discretization)}}                                 & \multicolumn{3}{c|}{\textbf{Accuracy (\%)}}                          \\ \cline{2-4} 
                                                                & \textbf{M7} & \textbf{M8} & \textbf{M9} \\ \hline
\textbf{\begin{tabular}[c]{@{}c@{}}Binary Class\end{tabular}} & 96.25 $\pm{0.23} $               & 96.01  $\pm{0.29} $           & 95.87  $\pm{0.32} $              \\ \hline
\textbf{\begin{tabular}[c]{@{}c@{}}Three Class\end{tabular}}  & 86.91   $\pm{0.31} $             & 86.47   $\pm{0.35} $          & 85.83  $\pm{0.39} $              \\ \hline
\end{tabular}
\end{table}
\begin{table}[h!]\setlength{\extrarowheight}{3pt}\caption{Adversarial Attack on Models Trained Using discretization-Based Transformed COVID-19 Chest X-Ray Images}\label{Attacks_DT}
~~\begin{tabular}{|c|c|c|c|c|c|c|}
\hline
\multirow{2}{*}{\textbf{Attack}} & \multicolumn{3}{c|}{\textbf{Accuracy (Binary Class)}}     & \multicolumn{3}{c|}{\textbf{Accuracy (Three Class)}}     \\ \cline{2-7} 
                                 & \textbf{M7} & \textbf{M8} & \textbf{M9} & \textbf{M7} & \textbf{M8} & \textbf{M9} \\ \hline
\textbf{PGD}                     & 96.10              & 95.97           & 95.51              & 86.53              & 86.12           & 85.41              \\ \hline
\textbf{PGD-r}                   & 96.07              & 95.91           & 95.49              & 86.51              & 86.10           & 85.39              \\ \hline
\textbf{FGSM}                    & 95.86              & 95.80           & 95.31              & 86.12              & 86.09           & 85.21              \\ \hline
\textbf{CW}                      & 95.81              & 95.79           & 95.29              & 86.10              & 85.91           & 85.17              \\ \hline
\textbf{Deep Fool}               & 95.77              & 95.61           & 95.18              & 85.97              & 85.89           & 85.12              \\ \hline
\textbf{BIM}                     & 95.70              & 95.47           & 94.97              & 85.91              & 85.81           & 84.96              \\ \hline
\end{tabular}
\end{table}
 For the standard discretization-based transformation, each time pixel value is divided by L and floor value is computed to know the interval to which the pixel value belongs. For an example, if value of the L is 32, then the total number of intervals is equal to $7$ when pixels value has a range between 0 to 255. In this study, we set the value of L as 32. The normal discretization is a hard assignment of intervals, and the proposed FUIT technique is a soft assignment of intervals. The model M7 shows the highest mean accuracy of 96.25\% and 86.91\% for the binary and three class classification, respectively. The models M7, M8, and M9 show less accuracy than the models M4, M5, and M6. The soft assignment of intervals using the FUIT technique is capable of dealing with uncertainty occurs during the assignment of the pixels into the intervals, thus resulting in higher accuracy of the model M4 compared to the model M7. Table \ref{Attacks_DT} shows the accuracy of the models M7, M8, and M9 for binary and three class classification in the presence of the six adversarial attacks. The models trained on images transformed using the typical discretization approach also show high accuracy to prevent the deep COVID-19 model against the adversarial attacks but show less accuracy as compared to models trained using the FUIT transformed images. It is clear from the Table \ref{Attacks_FUIT} and Table \ref{Attacks_DT} that the models trained using the FUIT transformed images are more secure against the adversarial attacks while diagnosis of the COVID-19 cases. All the nine models show high classification accuracy for the binary class classification and less accuracy for three class-classifications. The COVID-19 cases share similar symptoms with the pneumonia cases, which results in less classification accuracy in the case of the three-class classification.
\subsection{\textbf{Results on CT Image Dataset}}
The proposed model is also evaluated on second available CT Scan Image Dataset \cite{zhao2020covid} for the diagnosis of COVID-19. The dataset contains 398 images for normal patients and 399 images for the patients suffer from COVID-19. 
\begin{figure*}[h!] 
	\centering

	\begin{subfigure}[h]{.4\textwidth}
		\includegraphics[width=6.5 cm,height=4 cm]{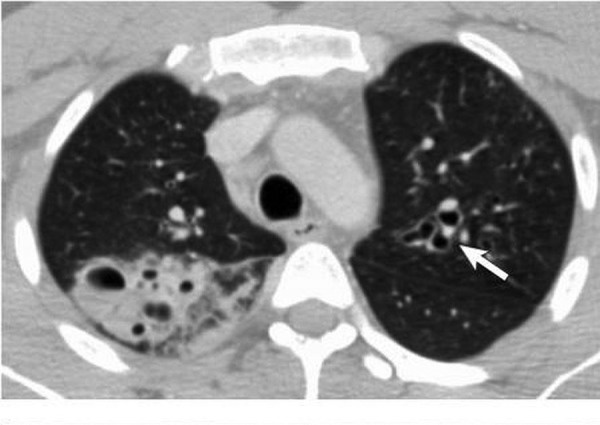}
		\caption{CT Image of Normal Patient-1}
		\label{CTN1}
	\end{subfigure} %\qquad
	\begin{subfigure}[h]{.4\textwidth}
		\includegraphics[width=6.5 cm,height=4 cm]{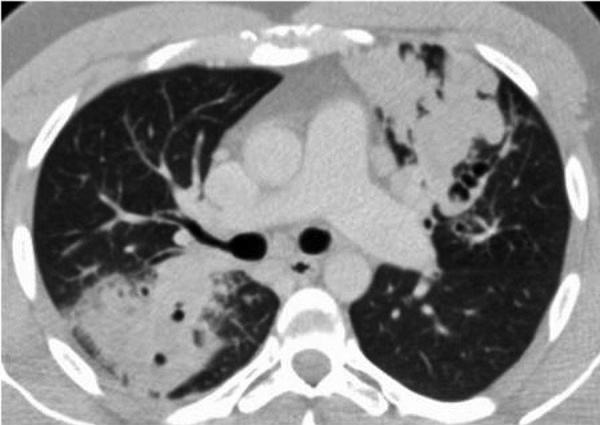}
		\caption{CT Image of Normal Patient-2}
		\label{CTN2}
	\end{subfigure} %\qquad
	\begin{subfigure}[h]{.4\textwidth}
		\includegraphics[width=6.5 cm,height=4 cm]{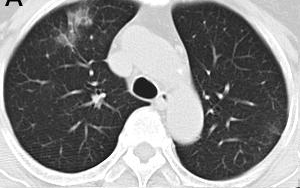}
		\caption{CT Image of COVID-19 Patient-1}
		\label{CTCovid-1}
	\end{subfigure}
	\begin{subfigure}[h]{.4\textwidth}
		\includegraphics[width=6.5 cm,height=4 cm]{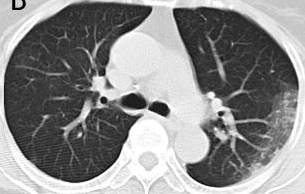}
		\caption{CT Image of COVID-19 Patient-2}
		\label{CTCovid-2}
	\end{subfigure} %\qquad
	\caption{Sample Images of Normal, COVID-19 Cases From CT Image Dataset \cite{cohen2020covid}}\label{CT_Dataset}
\end{figure*}
\begin{figure*}[!] 
	\centering
	\begin{subfigure}[!]{.4\textwidth}
		\includegraphics[width=6.5 cm,height=4 cm]{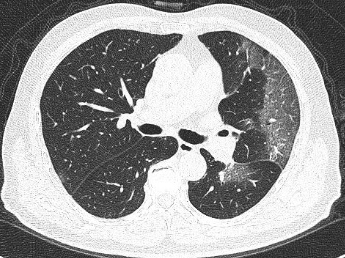}
		\caption{}
		\label{COVID_Class}
	\end{subfigure} %\qquad
	\begin{subfigure}[!]{.4\textwidth}
		\includegraphics[width=6.5 cm,height=4 cm]{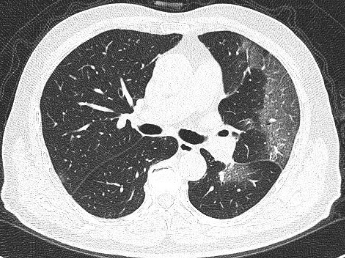}
		\caption{}
		\label{Adv_COVID_Class}
		\end{subfigure} 
		\caption{Clean and Adversarial CT Image of COVID-19 Class Misclassified By Model M10 Under PGD Attack, (a) Clean COVID-19 CT Image Classified by Model M10 with $Prob (COVID19 )= 0.85$ , $Prob (Normal)= 0.15 $, (b) Adversarial COVID-19 CT Image Misclassified with 
		$Prob(COVID19)=0.41 $ , $Prob(Normal)=0.59$} \label{Attack_Effect}
		
\end{figure*}
Total of nine models (M10-M18) are trained by initialize the weighted using the three pre-trained models Resnet-18 \cite{he2016deep}, VGG-16 \cite{simonyan2014very} and GoogLeNet \cite{szegedy2015going}. Table \ref{CTModels} shows a description of developed models to evaluate the performance of the proposed method on CT Image Dataset. Classification of COVID-19 cases in CT images is a problem of binary classification. The evaluation is performed by applying 5-fold cross-validation technique. Table \ref{CT_Accuracy} shows the accuracy of nine models developed to classify the COVID-19 cases.
Fig.(\ref{CT_Dataset}) shows sample images belong to normal and COVID-19 cases in the CT image dataset. 
\begin{table}[h!]\setlength{\extrarowheight}{3pt}\caption{Description of Models Developed for CT Image Dataset} \label{CTModels}\scalebox{0.90}{
\begin{tabular}{|c|c|c|c|}
\hline
\textbf{Model} & \textbf{Dataset} & \textbf{Image Type}   & \textbf{Pre-Trained Model} \\ \hline
M10            & CT Image          & Clean Image                    & Resnet-18                  \\ \hline
M11            & CT Image          & Clean Image                    & VGG-16                     \\ \hline
M12            & CT Image         & Clean Image                    & GoogLeNet                  \\ \hline
M13            & CT Image          & FUIT Transformed Image         & Resnet-18                  \\ \hline
M14            & CT Image          & FUIT Transformed Image         & VGG-16                     \\ \hline
M15            & CT Image          & FUIT Transformed Image         & GoogLeNet                  \\ \hline
M16            & CT Image         & Discretization Transformed Image & Resnet-18                  \\ \hline
M17            & CT Image          & Discretization Transformed Image & VGG-16                     \\ \hline
M18            & CT Image         & Discretization Transformed Image & GoogLeNet                  \\ \hline
\end{tabular}}
\end{table}

\begin{table}[ht!]\setlength{\extrarowheight}{3pt}\caption{Performance of Different Models to Classify Images of CT Image Datset}\label{CT_Accuracy}
~~~~~~~~~~~~~~~~~~~~~~~~~~~~~\begin{tabular}{|c|c|}
\hline
\textbf{Model} & \textbf{Accuracy (\%)} \\ \hline
M10            & 89.19  $\pm{0.13}$              \\ \hline
M11            & 89.12   $\pm{0.18}  $             \\ \hline
M12            & 88.37    $\pm{0.22}$              \\ \hline
M13            & 88.31    $ \pm{0.17} $            \\ \hline
M14            & 88.25  $\pm{0.24} $               \\ \hline
M15            & 88.07    $\pm{0.19} $             \\ \hline
M16            & 88.03    $\pm{0.12} $             \\ \hline
M17            & 87.98    $ \pm{0.11}$             \\ \hline
M18            & 87.77    $\pm{0.21} $             \\ \hline
\end{tabular}
\end{table}

 The model M10 achieves the highest mean accuracy of  89.19\%. The model M18 shows the lowest mean accuracy of 87.77\%. The developed models are tested to classify the COVID-19 cases in the presence of six adversarial attacks from the Table \ref{AttacksUsed}. The value of L is set as 32 for models M16, M17, and M18.  
 Table \ref{CT_ADVAttack} shows the accuracy of models to classify the COVID-19 cases under the six attacks. The models M10, M11, and M12 show degradation in classification accuracy when tested against the adversarial CT images. The BIM attack is again the most successful attack that drops the accuracy of model M10 from 89.19\% to 8.12\%. The drop in classification accuracy for model M11 and M12 is 81.07\% and 79.86\%, respectively. The highest classification accuracy of models M10, M11, and M12 under PGD attack are 48.19\%, 47.81\%, and 46.65\%, respectively. However, the models M13, M14, and M15 show the accuracy of 86.12\% and 86.01\% and 85.97\% respectively under the PGD attack. 

 	\begin{table}[h!]\setlength{\extrarowheight}{5pt}\caption{Performance of Different Models Under Six Different Adversarial Attacks}\label{CT_ADVAttack}\scalebox{0.77}{
\begin{tabular}{|c|c|c|c|c|c|c|c|c|c|}
\hline
\multirow{3}{*}{\textbf{Attack}} & \multicolumn{9}{c|}{\textbf{Accuracy (\%)}}                                                                                             \\ \cline{2-10} 
                                 & \multicolumn{9}{c|}{\textbf{Models}}                                                                                                    \\ \cline{2-10} 
                                 & \textbf{M10}  & \textbf{M11}  & \textbf{M12}  & \textbf{M13} & \textbf{M14} & \textbf{M15} & \textbf{M16} & \textbf{M17} & \textbf{M18} \\ \hline
\textbf{PGD}                     & 48.19         & 47.81         & 46.65         & 86.12        & 86.01        & 85.97        & 86.02        & 85.58        & 85.37        \\ \hline
\textbf{PGD-r}                   & 47.52         & 47.39         & 46.01         & 86.07        & 86.03        & 85.98        & 85.87        & 85.81        & 85.62        \\ \hline
\textbf{FGSM}                    & 43.71         & 43.59         & 41.92         & 86.03        & 86.01        & 85.93        & 85.81        & 85.77        & 85.61        \\ \hline
\textbf{CW}                      & 9.89          & 9.71          & 9.59          & 85.97        & 85.87        & 85.91        & 85.80        & 85.72        & 85.57        \\ \hline
\textbf{Deep Fool}               & 12.19         & 12.01         & 11.86         & 85.92        & 85.83        & 85.78        & 85.71        & 85.70        & 85.56        \\ \hline
\textbf{BIM}                     & \textbf{8.12} & \textbf{8.05} & \textbf{7.91} & 85.21        & 85.19        & 85.01        & 85.02        & 85.01        & 85.01        \\ \hline
\end{tabular}}
\end{table}
 \par It is clear from the Table \ref{CT_ADVAttack} that the models M13, M14 and M15 are more secure towards the six adversarial attacks and maintain the high accuracy to classify the COVID-19 cases. The accuracy of the models M13, M14 and M15 for the BIM attack are 85.21\%, 85.19\% and 85.01\% respectively. However the accuracy of the models M16, M17 and M18 are 85.02\%, 85.01\% and 85.01\%  respectively. The accuracy of the models M16, M17 and M18 are little less than the model M13, M14 and M15 which shows the FUIT technique efficiently deal with uncertainty created during the downsampling of the image pixels into an interval. The performance of the models M13, M14 and M15 as shown in the Table \ref{CT_ADVAttack} verifies that FUIT transformed images make the deep model more secure against the adversarial attacks and helpful to develop mode reliable deep models for the diagnosis of COVID-19 cases. \par
\begin{figure}[!]
	~~~\includegraphics[width=8 cm,height=5cm]{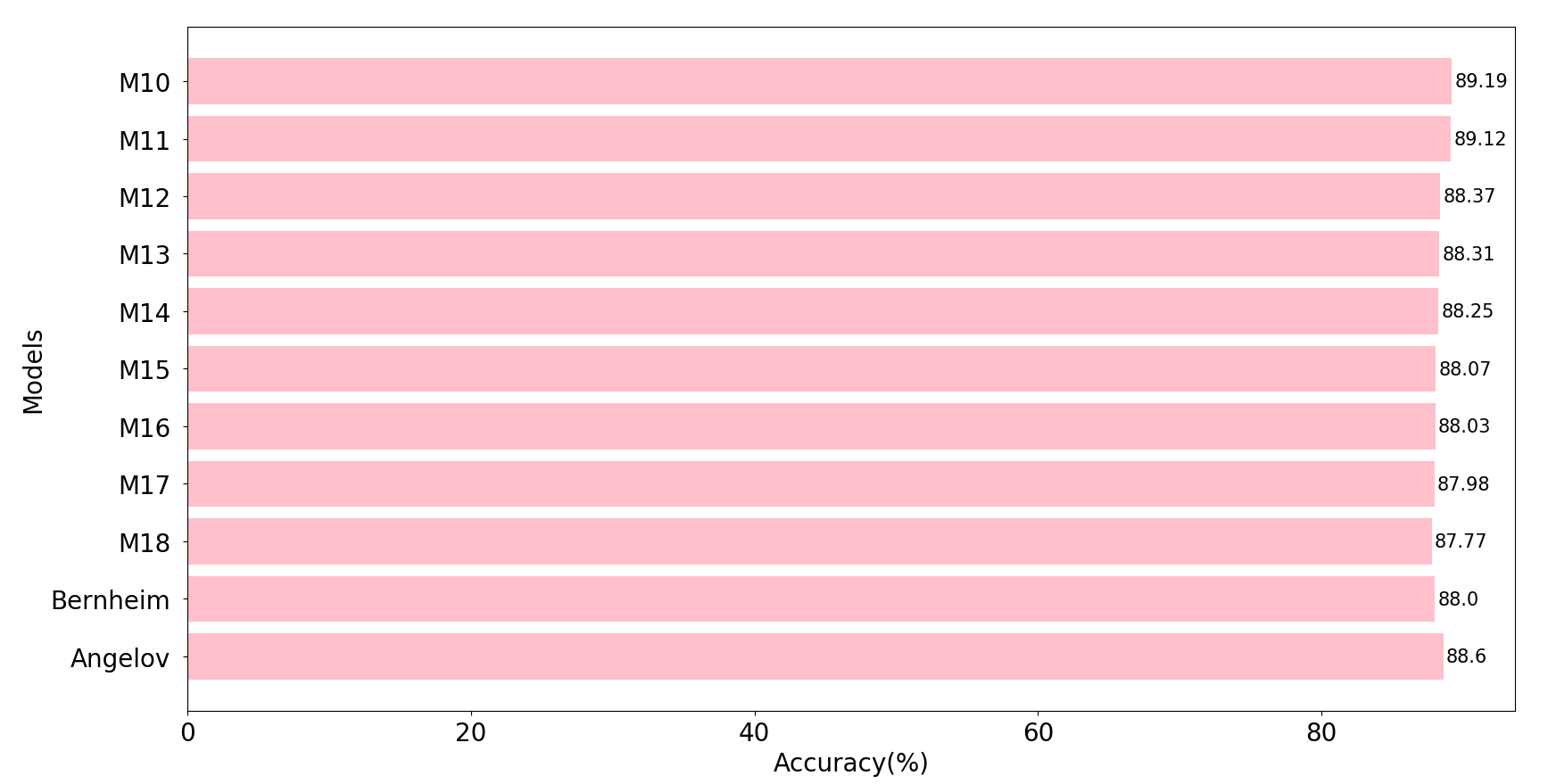}
	\caption{Comparison of Accuracy of Different Models for Binary Class Classification to Classify CT Image Dataset}
	\label{CT_Binary}
	\end{figure}

 Fig.(\ref{COVID_Class}) shows a clean COVID-19 CT image correctly classified as COVID-19 case with a class probability of 0.85 by the model M10. On the other hand when the model is attacked with PGD attack the model classifies the same image as a Normal case with a class probability of 0.59. Fig.(\ref{Adv_COVID_Class}) shows the adversarial image of the COVID-19 case (adversarial image of the image shown in Fig.(\ref{COVID_Class}) ). In presence of PGD attack the class probability of COVID-19 decreases to 0.41. The difference between the clean COVID-19 CT image and adversarial COVID-19 CT is visually unrecognizable by humans but well recognized by the deep model. Fig.(\ref{Attack_Effect}) shows comparison of these two images when classified by model M10 in presence of the PGD attack. Table \ref{CT_SOTA} shows comparison of the proposed model with SOTA methods to classify the COVID-19 cases using the CT images. Fig.(\ref{CT_Binary}) shows comparison of accuracy of different models to classify the COVID-19 cases in the CT image dataset.

% \usepackage{multirow}
\begin{table}[!]\setlength{\extrarowheight}{3pt}\caption{Comparison of Proposed Model With SOTA Methods to Classify CT Images}
\label{CT_SOTA}
~~~~~~~~~~~~~~~~~~~~~\begin{tabular}{|c|c|}
\hline
\textbf{Model} & \textbf{Accuracy(\%)} \\ \hline
M10            & 89.19              \\ \hline
M11            & 89.12               \\ \hline
M12            & 88.37                 \\ \hline
M13            & 88.31               \\ \hline
M14            & 88.25                \\ \hline
M15            & 88.07                 \\ \hline
M16            & 88.03                 \\ \hline
M17            & 87.98                 \\ \hline
M18            & 87.77                \\ \hline
 Bernheim et al. \cite{bernheim2020chest}              &    88                   \\ \hline
 Angelov et al. \cite{angelov2020explainable}              &   88.60                    \\ \hline
\end{tabular}
\end{table}
\section{Conclusion}
In this paper, we presented a novel fuzzy unique image transformation (FUIT) technique as a pre-processing step that prevents the COVID-19 deep model against the adversarial attacks. The FUIT technique downsamples the image pixels into an interval by using the created fuzzy sets. The FUIT technique prevents an increase in the variance of the number of unique pixels of the given image. This results in an equal number of unique pixels values in the clean and adversarial images. The deep model trained using the FUIT transformed images shows robust and secure performance against the adversarial attacks. The experiment and results on two available COVID-19 diagnosis datasets validate that the model maintains high accuracy to classify the COVID-19 cases in various non-targeted adversarial attacks. Moreover, the proposed model trains more secure deep models that prevent the COVID-19 deep model from the adversarial attacks. 
\par The study is performed using datasets with significantly fewer images, which could be one drawback of this study. In future, the models will be trained on more images collected from other publicly available repositories and nearby local hospitals. Besides, an inspection of the FUIT to develop the deep models to classify the images received from various research domains can be considered a natural extension of this study.

% or


%\appendices

%\section{}

%\section*{Acknowledgment}


\ifCLASSOPTIONcaptionsoff
  \newpage
\fi



%\bibliographystyle{IEEEtran}
%\clearpage
\FloatBarrier
\bibliographystyle{IEEEtran}
\bibliography{IEEEabrv,FUIT}


%\end{IEEEbiography}


%\newpage



%\vfill


\end{document}